\documentclass[a4paper]{amsart}
\usepackage{epic,eepic}
\usepackage{epsfig,cite,indentfirst}
\usepackage{multicol,boxedminipage}
\usepackage{varioref} 

\begin{document}

\title[DWBC trigonometric Felderhof model]
{On the trigonometric Felderhof model with domain wall boundary conditions}

\author{A Caradoc, O Foda, M Wheeler and M Zuparic}

\address{Department of Mathematics and Statistics,
         University of Melbourne, 
         Parkville, Victoria 3010, Australia.}

%\email{foda, 
%    dondrei, 
%   mwheeler,
%       mzup@ms.unimelb.edu.au}

\keywords{Trigonometric Felderhof model,
          domain wall boundary conditions} 
\subjclass[2000]{Primary 82B20, 82B23}
\date{}

\newcommand{\field}[1]{\mathbb{#1}}
\newcommand{\CC}{\field{C}}
\newcommand{\NN}{\field{N}}
\newcommand{\ZZ}{\field{Z}}

\newcommand{\B}{{\mathcal B}}
\newcommand{\D}{{\mathcal D}}
\newcommand{\M}{{\mathcal M}}
\newcommand{\R}{{\mathcal R}}
\newcommand{\T}{{\mathcal T}}
\newcommand{\U}{{\mathcal U}}
\newcommand{\Y}{{\mathcal Y}}
\newcommand{\Z}{{\mathcal Z}}

\newcommand{\tT}{\widetilde{\mathcal T}}
\newcommand{\tU}{\widetilde{\mathcal U}}

\renewcommand{\P}{{\mathcal P}}

\begin{abstract}
We consider the trigonometric Felderhof model, of free fermions in 
an external field, on a finite lattice with domain wall boundary 
conditions. The vertex weights are functions of rapidities and 
external fields. 

We obtain a determinant expression for the partition 
function in the special case where the dependence on the rapidities 
is eliminated, but for general external field variables. This
determinant can be evaluated in product form. In the homogeneous 
limit, it is proportional to a 2-Toda $\tau$ function. 

Next, we use the algebraic Bethe ansatz factorized basis to obtain 
a product expression for the partition function in the 
general case with dependence on all variables. 
\end{abstract}

\maketitle

\newtheorem{ca}{Figure}
\newtheorem{corollary}{Corollary}
\newtheorem{definition}{Definition}
\newtheorem{example}{Example}
\newtheorem{lemma}{Lemma}
\newtheorem{notation}{Notation}
\newtheorem{proposition}{Proposition}
\newtheorem{remark}{Remark}
\newtheorem{theorem}{Theorem}

\def\ll{\left\lgroup}
\def\rr{\right\rgroup}

\newcommand{\Proof}{\medskip\noindent {\it Proof: }}
\def\proofend{\ensuremath{\square}}

\def\no{\nonumber}

\def\union{\mathop{\bigcup}}
\def\vac{|\mbox{vac}\rangle}
\def\cav{\langle\mbox{vac}|}

\def\lprod{\mathop{\prod{\mkern-29.5mu}{\mathbf\longleftarrow}}}
\def\rprod{\mathop{\prod{\mkern-28.0mu}{\mathbf\longrightarrow}}}

\def\r{\rangle}
\def\l{\langle}

\def\a{\alpha}
\def\b{\beta}
\def\g{\gamma}
\def\d{\delta}
\def\e{\epsilon}
\def\l{\lambda}

\def\s{\sigma}

\def\eps{\varepsilon}
\def\hb{\hat\beta}

\def\tg{\operatorname{tg}}
\def\ctg{\operatorname{ctg}}
\def\sh{\operatorname{sh}}
\def\ch{\operatorname{ch}}
\def\cth{\operatorname{cth}}
\def\th{\operatorname{th}}

\def\tla{\tilde{\lambda}}
\def\tmu{\tilde{\mu}}

\def\sul{\sum\limits}
\def\pl{\prod\limits}

\def\pd #1{\frac{\partial}{\partial #1}}
\def\const{{\rm const}}
\def\argum{\{\mu_j\},\{\la_k\}} 
\def\umarg{\{\la_k\},\{\mu_j\}} 

\def\prodmu #1{\prod\limits_{j #1 k} \sinh(\mu_k-\mu_j)}
\def\prodla #1{\prod\limits_{j #1 k} \sinh(\lambda_k-\lambda_j)}

\newcommand{\bl}[1]{\makebox[#1em]{}}

\def\tr{\operatorname{tr}}
\def\Res{\operatorname{Res}}
\def\det{\operatorname{det}}

\newcommand{\boldN}{\boldsymbol{N}}
\newcommand{\bra}[1]{\langle\,#1\,|}
\newcommand{\ket}[1]{|\,#1\,\rangle}
\newcommand{\bracket}[1]{\langle\,#1\,\rangle}
\newcommand{\infinity}{\infty}

\renewcommand{\labelenumi}{\S\theenumi.}

\let\up=\uparrow
\let\down=\downarrow
\let\tend=\rightarrow
\hyphenation{boson-ic
             ferm-ion-ic
             para-ferm-ion-ic
             two-dim-ension-al
             two-dim-ension-al
             rep-resent-ative
             par-tition}

\renewcommand{\mod}{\textup{mod}\,}
\newcommand{\wt}{\text{wt}\,}

\hyphenation{And-rews
             Gor-don
             boson-ic
             ferm-ion-ic
             para-ferm-ion-ic
             two-dim-ension-al
             two-dim-ension-al}

\setcounter{section}{-1}

\section{Introduction}

In \cite{korepin}, Korepin introduced domain wall boundary conditions 
for the six vertex model on a finite square lattice, and proposed 
recursion relations that determine the corresponding domain wall 
partition function. In \cite{izergin}, Izergin obtained a determinant 
solution of Korepin's recursion relations. At the free fermion point, 
the six vertex domain wall partition function can be evaluated 
explicitly in product form \cite{bpz}. In the homogeneous limit, 
it is proportional to a 1-Toda $\tau$ function \cite{sogo}. 

In this work, we look for analogous results in the context of the 
trigonometric limit of Felderhof's model \cite{felderhof}, which is 
a vertex model of free fermions in an external field. In section 
{\bf 1}, we recall the definition of the model in the parametrization 
of Deguchi and Akutsu \cite{da}, and formulate it on an $N\times N$ 
lattice. There are four sets of complex variables:
horizontal and vertical rapidities $\{u_i, v_j\}$, and 
horizontal and vertical external field 
variables $\{\alpha_i, \beta_j\}$, where 
$\{i, j\} \in \{1, 2,\ldots, N\}$\footnote{In \cite{da}, the 
{\it external field variables} are referred to as {\it colour 
variables}.}.
The weight $w_{ij}$ of the vertex $v_{ij}$ at the intersection
of the $i$-th horizontal line and $j$-th vertical line depends 
on the difference of the rapidities, $u_i - v_j$, but depends on 
the external fields, $\alpha_i$ and $\beta_j$, separately. 

In section {\bf 2}, we impose domain wall boundary conditions and 
obtain an Izergin-type determinant expression for the domain wall 
partition function, under the restriction that the difference of 
any two rapidity variables is a multiple 
of $2 \pi \sqrt{-1}$, but for general $\{\alpha_i, \beta_j\}$. This 
expression can be evaluated in product form. In the homogeneous 
limit, it is proportional to a 2-Toda $\tau$ function 
\cite{JM, hirota-book}. 

In section {\bf 3}, we use the factorized basis of the algebraic 
Bethe ansatz, \cite{ms, kmt}, to obtain a product expression for 
the domain wall partition function for general 
$\{u_i, v_j\}$ and $\{\alpha_i, \beta_j\}$.

\subsection{Abbreviations}
In the rest of this paper, {\it DWBC} stands for {\it `domain wall 
boundary conditions'} and 
{\it DWPF} stands for {\it `domain wall partition function'}.
$Z^{N\times N}_{TF}$ is the DWPF of the trigonometric Felderhof 
model on an $N\times N$ lattice.
$Z^{N\times N}_{6V}$ is the DWPF of the six vertex 
model on an $N\times N$ lattice. 

$Z^{N\times N}_{TF, res}$ is $Z^{N\times N}_{TF}$ with restrictions 
on the rapidities as in Equation {\bf \ref{restrictions}}.
$Z^{N\times N}_{TF, res, hom}$ is the homogeneous version of
$Z^{N\times N}_{TF, res}$ with 
all horizontal external field variables equal, and 
all vertical   external field variables equal.

\section{The trigonometric Felderhof model}

%\begin{multicols}{2}

\subsection{The lattice} We work on a square lattice consisting 
of $N$ horizontal and $N$ vertical lines. We label the horizontal 
lines from top to bottom, and the vertical lines from left to right. 

We assign the {\it i}-th horizontal line an orientation from left 
to right, a complex rapidity variable $u_i$ and a complex external 
field variable $\alpha_i$. 
We assign the {\it j}-th vertical line an orientation from bottom 
to top, a complex rapidity variable $v_j$ and a complex external
field variable $\beta_j$. 

%FIG-01
%
\begin{center}
\begin{minipage}{4.9in}
\setlength{\unitlength}{0.0008cm}
\begin{picture}(4800, 6000)(-1000, -0300)
% negative x shifts the figure to the right 
% negative y shifts the figure up
\thicklines
\path(2400,5400)(2400,1800)
\path(3000,5400)(3000,1800)
\path(3600,5400)(3600,1800)
\path(4200,5400)(4200,1800)
\path(4800,5400)(4800,1800)
\path(1800,4800)(5400,4800)
\path(1800,4200)(5400,4200)
\path(1800,3600)(5400,3600)
\path(1800,3000)(5400,3000)
\path(1800,2400)(5400,2400)
\path(0600,4254)(1200,4254)
\path(2400,654)(2400,1254)
\path(3000,654)(3000,1254)
\path(3600,654)(3600,1254)
\path(4200,654)(4200,1254)
\path(4800,654)(4800,1254)
\path(600,2454)(1200,2454)
\path(600,3054)(1200,3054)
\path(600,3654)(1200,3654)
\path(600,4854)(1200,4854)
\whiten\path(2490,894)(2400,1254)(2310,894)(2490,894)
\whiten\path(3090,894)(3000,1254)(2910,894)(3090,894)
\whiten\path(3690,894)(3600,1254)(3510,894)(3690,894)
\whiten\path(4290,894)(4200,1254)(4110,894)(4290,894)
\whiten\path(4890,894)(4800,1254)(4710,894)(4890,894)
\whiten\path(840,2364)(1200,2454)(840,2544)(840,2364)
\whiten\path(840,2964)(1200,3054)(840,3144)(840,2964)
\whiten\path(840,3564)(1200,3654)(840,3744)(840,3564)
\whiten\path(840,4164)(1200,4254)(840,4344)(840,4164)
\whiten\path(840,4764)(1200,4854)(840,4944)(840,4764)
%
%the -1000 shifts things to the left 

\put(-1000,4854){$u_1, \alpha_1$}
\put(-1000,4254){$u_2, \alpha_2$}
\put(-1000,2454){$u_N, \alpha_N$}
\put(2300, 0250){$v_1$}
\put(2300,-0250){$\beta_1$}
\put(2900, 0250){$v_2$}
\put(2900,-0250){$\beta_2$}
\put(4700, 0250){$v_N$}
\put(4700,-0250){$\beta_N$}
\end{picture}
\begin{ca} 
\label{lattice}
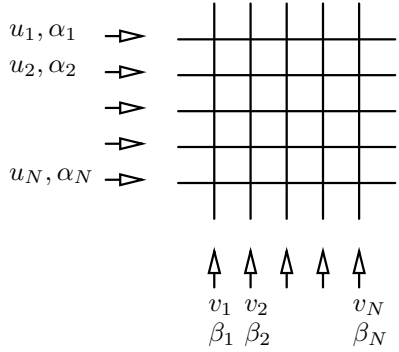
An $N\times N$ square lattice, with oriented lines. To each line, 
we attach two complex variables, a rapidity and an external field. 
\end{ca}

\end{minipage}
\end{center}

%\end{multicols}

\subsection{Vertices} Each line intersects with $N$ other 
lines. 
A line segment between two intersections is a bond. To each bond, 
we assign a state variable, namely an arrow that points along 
the orientation of that line segment or against it. 
The intersection of the $i$-th horizontal line and the $j$-th 
vertical line, together with the four bonds adjacent to it, and 
the set of arrows on these bonds, is a vertex $v_{ij}$. 

%FIG-02

\begin{center}
\begin{minipage}{4.3in}

\setlength{\unitlength}{0.0008cm}
\begin{picture}(10000,7500)(-1000, 0)
\thicklines

\blacken\path(10162,2310)(10522,2400)(10162,2490)(10162,2310)
\blacken\path(01282,2490)(00922,2400)(01282,2310)(01282,2490)
\blacken\path(01462,5610)(01822,5700)(01462,5790)(01462,5610)
\blacken\path(01732,1860)(01822,1500)(01912,1860)(01732,1860)
\blacken\path(01732,2685)(01822,2325)(01912,2685)(01732,2685)
\blacken\path(01912,5340)(01822,5700)(01732,5340)(01912,5340)
\blacken\path(01912,6240)(01822,6600)(01732,6240)(01912,6240)
\blacken\path(02182,2490)(01822,2400)(02182,2310)(02182,2490)
\blacken\path(02362,5610)(02722,5700)(02362,5790)(02362,5610)
\blacken\path(05182,5790)(04822,5700)(05182,5610)(05182,5790)
\blacken\path(05362,2310)(05722,2400)(05362,2490)(05362,2310)
\blacken\path(05632,1860)(05722,1500)(05812,1860)(05632,1860)
\blacken\path(05632,2760)(05722,2400)(05812,2760)(05632,2760)
\blacken\path(05812,5340)(05722,5700)(05632,5340)(05812,5340)
\blacken\path(05812,6240)(05722,6600)(05632,6240)(05812,6240)
\blacken\path(06082,5790)(05722,5700)(06082,5610)(06082,5790)
\blacken\path(06262,2310)(06622,2400)(06262,2490)(06262,2310)
\blacken\path(09082,2490)(08722,2400)(09082,2310)(09082,2490)
\blacken\path(09262,5610)(09622,5700)(09262,5790)(09262,5610)
\blacken\path(09532,2760)(09622,2400)(09712,2760)(09532,2760)
\blacken\path(09532,5160)(09622,4800)(09712,5160)(09532,5160)
\blacken\path(09712,2040)(09622,2400)(09532,2040)(09712,2040)
\blacken\path(09712,6240)(09622,6600)(09532,6240)(09712,6240)
\blacken\path(09982,5790)(09622,5700)(09982,5610)(09982,5790)

\path(10222,5700)(09622,5700)
\path(01222,5700)(01822,5700)
\path(01522,2400)(00922,2400)
\path(01822,1500)(01822,2100)
\path(01822,2325)(01822,2925)
\path(01822,3300)(01822,1500)
\path(01822,3900)(01822,4500)
\path(01822,5700)(01822,5100)
\path(01822,0600)(01822,1200)
\path(01822,6600)(01822,4800)
\path(00022,2400)(00622,2400)
\path(00022,5700)(00622,5700)
\path(02422,2400)(01822,2400)
\path(03922,2400)(04522,2400)
\path(03922,5700)(04522,5700)
\path(04822,2400)(06622,2400)
\path(04822,5700)(06622,5700)
\path(05122,2400)(05722,2400)
\path(05422,5700)(04822,5700)
\path(05722,1500)(05722,2100)
\path(05722,2400)(05722,3000)
\path(05722,3300)(05722,1500)
\path(05722,3900)(05722,4500)
\path(05722,5700)(05722,5100)
\path(05722,0600)(05722,1200)
\path(05722,6600)(05722,4800)
\path(05722,6600)(05722,6000)
\path(06022,2400)(06622,2400)
\path(06322,5700)(05722,5700)
\path(07822,2400)(08422,2400)
\path(07822,5700)(08422,5700)
\path(08722,2400)(10522,2400)
\path(08722,5700)(10522,5700)
\path(09022,5700)(09622,5700)
\path(00922,2400)(02722,2400)
\path(00922,5700)(02722,5700)
\path(09322,2400)(08722,2400)
\path(09622,2400)(09622,1800)
\path(09622,2400)(09622,3000)
\path(09622,3300)(09622,1500)
\path(09622,3900)(09622,4500)
\path(09622,4800)(09622,5400)
\path(09622,0600)(09622,1200)
\path(09622,6600)(09622,4800)
\path(09622,6600)(09622,6000)
\path(09922,2400)(10522,2400)

\whiten\path(1912,4140)(1822,4500)(1732,4140)(1912,4140)
\whiten\path(1912,0840)(1822,1200)(1732,0840)(1912,0840)
\whiten\path(0262,2310)(0622,2400)(0262,2490)(0262,2310)
\whiten\path(0262,5610)(0622,5700)(0262,5790)(0262,5610)
\whiten\path(4162,2310)(4522,2400)(4162,2490)(4162,2310)
\whiten\path(4162,5610)(4522,5700)(4162,5790)(4162,5610)
\whiten\path(5812,4140)(5722,4500)(5632,4140)(5812,4140)
\whiten\path(5812,0840)(5722,1200)(5632,0840)(5812,0840)
\whiten\path(8062,2310)(8422,2400)(8062,2490)(8062,2310)
\whiten\path(8062,5610)(8422,5700)(8062,5790)(8062,5610)
\whiten\path(9712,4140)(9622,4500)(9532,4140)(9712,4140)
\whiten\path(9712,0840)(9622,1200)(9532,0840)(9712,0840)

\put(1000,3500){$a_2$}
\put(5000,3500){$b_1$}
\put(9000,3500){$c_1$}

\put(1000,0000){$a_1$}
\put(5000,0000){$b_2$}
\put(9000,0000){$c_2$}

\end{picture}

\begin{ca}
\label{vertices}
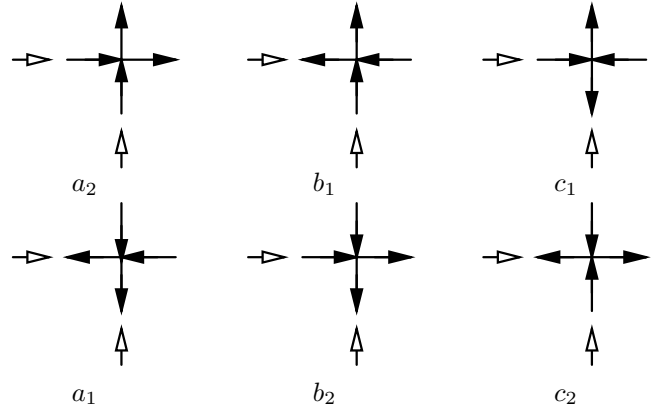
The non-zero weight vertices of the trigonometric Felderhof
model. The black arrows indicate state variables. The white 
indicate line orientation. Notice which vertex is $a_1$ and 
which is $a_2$.
\end{ca}

\end{minipage}
\end{center}

\subsection{Weights} To each vertex $v_{ij}$, we assign 
a weight $w_{ij}$, that depends on {\bf 1.} The orientations 
of the four arrows on the bonds of that vertex, {\bf 2.} The 
difference of rapidity variables flowing through the vertex, 
and {\bf 3.} The two external field variables flowing through 
the vertex. To satisfy the Yang-Baxter equations, only the six 
vertices shown in Figure {\bf \ref{vertices}} have non-zero 
weights \cite{da}. In the parametrization of \cite{da}, the 
non-zero weights are 

\begin{alignat}{3}
 a_{1}(\alpha_i, \beta_j, u_i, v_j) 
 &= 1 - \alpha_i \beta_j e^{u_i - v_j},
 &b_{1}(\alpha_i, \beta_j, u_i, v_j) 
  = \alpha_i - \beta_j   e^{u_i - v_j}, 
       \\
 a_{2}(\alpha_i, \beta_j, u_i, v_j) 
 &= e^{u_i - v_j} - \alpha_i \beta_j, 
 &b_{2}(\alpha_i, \beta_j, u_i, v_j) 
  = \beta_j  - \alpha_i e^{u_i - v_j},  
    \no \\
c_{1}(\alpha_i, \beta_j, u_i, v_j)  
&= \sqrt{1- \alpha_i^2} \sqrt{1- \beta_j^2} \ e^{u_i - v_j},
      \no \\
c_{2}(\alpha_i, \beta_j, u_i, v_j)  
&= \sqrt{1- \alpha_i^2} \sqrt{1- \beta_j^2}
    \no &
\label{general-weights}
\end{alignat}

In the sequel, we will drop the dependence on the variables, when 
that is clear from the indices. Unlike the six vertex model,  the 
vertex weights of the trigonometric Felderhof model are not 
invariant under reversing the directions of all the arrows.

\subsection{DWBC}

As in the six vertex model, the DWBC are such that all arrows on the left 
and right boundaries point inwards, and all arrows on the upper and lower 
boundaries point outwards. 

%FIG-03

\begin{center}
\begin{minipage}{4.9in}
\setlength{\unitlength}{0.0008cm}
\begin{picture}(4800, 4800)(-1000, 1500)

% negative x shifts the figure to the right 
% negative y shifts the figure up

\thicklines

%vertical lines
\path(2400,5400)(2400,1800) %left  most vertical line
\path(3000,5400)(3000,1800)
\path(3600,5400)(3600,1800)
\path(4200,5400)(4200,1800)
\path(4800,5400)(4800,1800) %right most vertical line 

%horizontal lines
\path(1800,4800)(5400,4800) %highest vertical line
\path(1800,4200)(5400,4200)
\path(1800,3600)(5400,3600)
\path(1800,3000)(5400,3000)
\path(1800,2400)(5400,2400) %lowest horizontal line

%horizontally ordered arrow -- above
%                             tip
\blacken\path(2490,5060)(2400,5420)(2310,5060)(2490,5060)
\blacken\path(3090,5060)(3000,5420)(2910,5060)(3090,5060)
\blacken\path(3690,5060)(3600,5420)(3510,5060)(3690,5060)
\blacken\path(4290,5060)(4200,5420)(4110,5060)(4290,5060)
\blacken\path(4890,5060)(4800,5420)(4710,5060)(4890,5060)

%horizontally ordered arrow -- below
%                             tip 
\blacken\path(2490,2140)(2400,1780)(2310,2140)(2490,2140)
\blacken\path(3090,2140)(3000,1780)(2910,2140)(3090,2140)
\blacken\path(3690,2140)(3600,1780)(3510,2140)(3690,2140)
\blacken\path(4290,2140)(4200,1780)(4110,2140)(4290,2140)
\blacken\path(4890,2140)(4800,1780)(4710,2140)(4890,2140)

%vertically ordered arrows -- left
\blacken\path(2040,4710)(2400,4800)(2040,4890)(2040,4710)
\blacken\path(2040,4110)(2400,4200)(2040,4290)(2040,4110)
\blacken\path(2040,3510)(2400,3600)(2040,3690)(2040,3510)
\blacken\path(2040,2910)(2400,3000)(2040,3090)(2040,2910)
\blacken\path(2040,2310)(2400,2400)(2040,2490)(2040,2310)

%vertically ordered arrows -- right

%                        tip 

\blacken\path(5160,4710)(4800,4800)(5160,4890)(5160,4710)
\blacken\path(5160,4110)(4800,4200)(5160,4290)(5160,4110)
\blacken\path(5160,3510)(4800,3600)(5160,3690)(5160,3510)
\blacken\path(5160,2910)(4800,3000)(5160,3090)(5160,2910)
\blacken\path(5160,2310)(4800,2400)(5160,2490)(5160,2310)

\end{picture}
\begin{ca} 
\label{dwbc}
Domain wall boundary conditions
\end{ca}
\end{minipage}
\end{center}

%\end{multicols}

\subsection{DWPF} Given a 2-state vertex model, such as the six
vertex model or the trigonometric Felderhof model, the DWPF on 
an $N\times N$ lattice, $Z^{N\times N}_{DWBC}$, is defined as 
the sum over all weighted configurations that satisfy DWBC. The 
weight of each configuration is the product of the weights of 
the vertices 

\begin{equation}
Z^{N\times N}_{DWBC}
=
\sul_{{\rm config-} \atop{\rm urations}}
\ll 
\pl_{\rm vertices} w_{ij} 
\rr
\label{physical}
\end{equation}

It is also possible to define DWBC and DWPF's in vertex models 
with more state variables \cite{cfk, df, zz}. 

\section{A determinant form of the restricted DWPF}

\subsection{The Korepin-Izergin procedure} As a first step 
towards computing $Z^{N\times N}_{TF}$, we follow the 
Korepin-Izergin procedure: 
{\bf 1.} Specify a set of properties that fully determine 
$Z^{N\times N}_{TF}$, 
{\bf 2.} Conjecture a determinant expression for the required
$Z^{N\times N}_{TF}$, and 
{\bf 3.} Show that the conjectured expression satisfies the 
required properties. 

It turns out that it is not obvious how to follow the above 
procedure for general values of {\it all} variables.  
The reason is that an Izergin-type determinant solution is 
tightly related to the Korepin-type properties, one of which 
is that the DWPF is {\it symmetric} under permuting the variables 
on any two parallel lattice lines. 

In the six vertex model, and in models discussed in \cite{cfk, 
df, zz}, this condition is automatically satisfied because the 
vertex weights are invariant under reversing the directions 
of all arrows, so the two $a$-type vertices, which are involved 
in proving this symmetry, have the same weight. In the trigonometric 
Felderhof model, there is no such invariance for general values of 
all variables, and we need to impose restrictions on at least some 
of the variables. 

Our plan is to restrict the variables to a point where the 
Korepin-Izergin prescription works. We claim that there is 
no Izergin-type determinant expression for $Z^{N\times N}_{TF}$,  
for general values of rapidities and external fields.

\subsection{Restrictions} We require 

\begin{equation}
e^{u_{i1} - u_{i2}} = 
e^{v_{j1} - v_{j2}} = 
e^{u_{i } - v_{j }}   =
1
\label{restrictions}
\end{equation}

The restrictions in Equation {\bf \ref{restrictions}} are satisfied 
by choosing the difference between any two rapidity variables to be 
a multiple of $2 \pi \sqrt{-1}$, or equivalently, by simply setting 
all rapidities to zero. The external field variables remain free. 
By eliminating the dependence on the rapidities, the weights are now 
much simpler and can be written as

\begin{eqnarray}
a_{0, ij} &=& a_{1}(\alpha_i, \beta_j, 0, 0) = \phantom{-}  
              a_{2}(\alpha_i, \beta_j, 0, 0) = 1 - \alpha_i \beta_j,   
\\
b_{0, ij} &=& b_{1}(\alpha_i, \beta_j, 0, 0) =          -   
              b_{2}(\alpha_i, \beta_j, 0, 0) =     \alpha_i - \beta_j, 
	     \no  \\ 
c_{0, ij} &=& c_{1}(\alpha_i, \beta_j, 0, 0) = \phantom{-}  
              c_{2}(\alpha_i, \beta_j, 0, 0) = 
\sqrt{1 - \alpha_i^2} \sqrt{1 - \beta_j^2}
\no
\end{eqnarray}

Under these conditions, 
$Z^{N\times N}_{TF}$ becomes 
$Z^{N\times N}_{TF, res}$.

\subsection{Korepin-type properties}\label{properties}

$Z^{N\times N}_{TF, res}$ is fully determined by the following 
properties

\begin{enumerate}

%1
\item[{\bf 1.}] It is symmetric in the elements of each of the sets 
$\{\alpha_{}\}$, and $\{ \beta_{}\}$. 

%2
\item[{\bf 2.}] It is a polynomial of degree $(N-1)$ 
in $\alpha_i$, up to a factor of $\sqrt{1 - \alpha_i^2}$, 
where $1 \leq i \leq N$, and 
in $\beta_j$,  up to a factor of $\sqrt{1 -  \beta_j^2}$,
where $1 \leq j \leq N$.

%3
\item[{\bf 3.}] It satisfies the recursion relation 

\begin{equation}
\left.
Z^{N\times N}_{TF, res} \rule{0mm}{3mm} 
\right|_{\alpha_{m} = \beta_{n}} 
=
c_{0, mn}
\ll 
\prod_{\stackrel{i=1}{i\neq  m}}^{N} a_{0, in} 
\rr
\ll
\prod_{\stackrel{j=1}{j\neq  n}}^{N} a_{0, mj}   
\rr
Z^{(N-1)\times (N-1)}_{TF, res, (mn)} 
\label{recursion-relation}
\end{equation}

\noindent for any $m, n \in \{1, \ldots, N \}$.
The subscripts $(mn)$ in $Z^{(N-1)\times (N-1)}_{TF, res, (mn)}$
indicate the variables that are not present in 
the reduced partition function.

%4
\item[{\bf 4.}] It satisfies the initial condition 
$
Z^{1\times 1}_{TF, res} =
c_{0, 11}$.

\end{enumerate}

\subsection{Izergin-type determinant solution} 

Using the notation $\alpha_{[ij]} = \alpha_i - \alpha_j$, {\it etc}, 
the properties in subsection {\bf \ref{properties}} are satisfied by

\begin{boxedminipage}[c]{12cm}
\begin{multline}
Z^{N\times N}_{TF, res} 
= 
\frac{
\prod_{1 \leq i, j \leq N} a_{0, ij} b_{0, ij}
}
{
\prod_{1 \leq i < j \leq N} \alpha_{[ij]} \beta_{[ji]}
}
\, \, 
\det 
\ll 
\frac{c_{0, ij}}{a_{0, ij} b_{0, ij}} 
\rr_{1 \leq i, j \leq N}
=
\\
\frac{
\prod_{1 \leq i, j \leq N} 
( 1       -  \alpha_i \beta_j) 
(\alpha_i -  \beta_j         )
}
{
\prod_{1 \leq i < j \leq N} (\alpha_i -  \alpha_j) 
                            ( \beta_j  -  \beta_i)
} 
\ll
\prod_{1 \leq k \leq N} 
\sqrt{1 - \alpha_{k}^{2}}
\sqrt{1 -  \beta_{k}^{2}} 
\rr
\\
\times
\det 
\ll \frac{1}{(1 - \alpha_i \beta_j)(\alpha_i - \beta_j)} 
\rr_{1 \leq i, j \leq N} 
\label{determinant-form}
\end{multline}
\end{boxedminipage}

\subsection{Remarks on proof of Equation 
{\bf \ref{determinant-form}}}
The proof proceeds along the same lines as Izergin's proof, which is 
discussed in detail in the literature, including \cite{korepin-book, 
df} and references therein, it suffices to outline it. Since the 
four Korepin-properties in {\bf \ref{properties}} fully determine 
$Z_{TF, res}^{N\times N}$, all we need to do is to show that the 
right hand side of Equation {\bf \ref{determinant-form}} satisfies 
each of these properties. 

Properties {\bf 1} and {\bf 2} can be checked precisely the same
way as in the case of the six vertex model \cite{korepin-book, df}. 
Property {\bf 4} can be checked by inspection. Property {\bf 3}
can be checked as follows.

Expanding the determinant in the right hand side of Equation 
{\bf 6} along the first row, we obtain

\begin{equation}
Z^{N\times N}_{TF, res} 
= 
\sum_{i=1}^{N}
(-)^{i+1}
c_{0,1i}
\frac{
\ll
\prod_{j\not= i} a_{0, 1j} b_{0, 1j}
\rr
\ll
\prod_{j=2}^{N} a_{0,ji} b_{0,ji}
\rr
}
{ 
\prod_{j=2}^{N} \alpha_{[1j]} 
\prod_{j<i} \beta_{[ij]}
\prod_{j>i} \beta_{[ji]}}
Z^{(N-1)\times (N-1)}_{TF, res, (1i)
}
\label{row-expansion} 
\end{equation}

Because of the DWBC, the vertex at the upper right corner must be 
either $b_1$ or $c_1$. 
By choosing $\alpha_1=$$\beta_N$, we eliminate the possibility 
of a $b_1$ vertex, and restrict the allowed configurations as 
follows
{\bf 1.} The upper right corner is a 
$c_{1}(\alpha_1, \beta_N, 0, 0) = c_{0, 1N}$ vertex, 
{\bf 2.} The right most column, apart from the upper right 
vertex, is a set of 
$a_{1}(\alpha_i, \beta_N, 0, 0) = a_{0, iN}$ vertices, 
where $\{2 \leq i \leq N  \}$, and 
{\bf 3.} The top row, apart from the upper right vertex, 
is a set of $a_{2}(\alpha_1, \beta_j, 0, 0) = a_{0, 1j}$ 
vertices, where $\{1 \leq j \leq N-1\}$. 

Letting $\alpha_1=\beta_N$ in Equation {\bf \ref{row-expansion}},
we obtain 

\begin{multline}
\left.
Z^{N\times N}_{TF, res} 
\rule{0mm}{3mm} 
\right|_{\alpha_{1} = \beta_{N}} 
=
\\
(-)^{N+1}
c_{0, 1N}
\frac{
\ll
\prod_{j=1}^{N-1} a_{0, 1j} b_{0, 1j}
\rr
\ll
\prod_{j=2}^{N} a_{0,jN} b_{0,jN}
\rr
}
{ 
\prod_{j=2}^{N} \alpha_{[1j]} 
\prod_{j=1}^{N-1} \beta_{[Nj]}
}
Z^{(N-1)\times (N-1)}_{TF, res, (1N)}
\\
=
c_{0,1N}
\ll
\prod_{j=1}^{N-1} a_{0, 1j} 
\rr
\ll
\prod_{j=2}^{N} a_{0, jN} 
\rr
Z^{(N-1)\times (N-1)}_{TF, res, (1N)}
\end{multline}

\noindent which is the recursion relation of Equation
{\bf \ref{recursion-relation}}, as we expect. Thus the
determinant expression on the right hand side of Equation 
{\bf \ref{determinant-form}} satisfies all Korepin-properties.

\subsection{Further check on Equation {\bf \ref{determinant-form}}} 

The proof of Equation {\bf \ref{determinant-form}} outlined 
above made use of a recursion relation obtained by freezing 
the upper right corner. We could have also chosen to freeze 
the upper left corner. Let us check that Equation 
{\bf \ref{determinant-form}} satisfies that second 
recurrence relation as well.

Because of the DWBC, the vertex at the upper left corner must 
be either $a_2$ or $c_1$.
By choosing $\alpha_1 \beta_1=$$1$, we eliminate the possibility 
of an $a_2$ vertex, and restrict the allowed configurations as 
follows
{\bf 1.} The upper left corner is a
$c_{1}(\alpha_1, \beta_1, 0, 0) = c_{0, 11}$ vertex, 
{\bf 2.} The left most column, apart from the upper left vertex 
is a set of 
$b_{2}(\alpha_i, \beta_1, 0, 0) =  - b_{0, i1}$ vertices, where 
$\{2 \leq i \leq N  \}$, and 
{\bf 3.} The top row, apart from the upper left vertex, 
is a set of 
$b_{1}(\alpha_1, \beta_j, 0, 0) =    b_{0, 1j}$ vertices, 
where $\{2 \leq j \leq N  \}$. 

Letting 
$\alpha_1 \beta_1=1$ in Equation {\bf \ref{row-expansion}},
we obtain

\begin{multline}
\left.
Z^{N\times N}_{TF, res} \rule{0mm}{3mm} 
\right|_{\alpha_{1} \beta_{1}=1} 
=
c_{0,11}
\frac{
\ll
\prod_{j=2}^{N} a_{0, 1j} b_{0, 1j}
\rr
\ll
\prod_{j=2}^{N} a_{0,j1} b_{0,j1}
\rr
}
{ 
\prod_{j=2}^{N} \alpha_{[1j]} 
\prod_{j=2}^{N} \beta_{[j1]}
}
Z^{(N-1)\times (N-1)}_{TF, res, (11)}
\\
=
c_{0,11}
\ll
\prod_{j=2}^{N} b_{0, 1j} 
\rr
\ll
\prod_{j=2}^{N} (-b_{0,j1}) 
\rr
Z^{(N-1)\times (N-1)}_{TF, res, (11)}
\label{another-recursion-relation}
\end{multline}

\noindent which is what we expect. 
Equation {\bf \ref{another-recursion-relation}} is a second 
recursion relation for $Z^{N\times N}_{TF, res}$, and provides 
an independent check of Equation {\bf \ref{determinant-form}}.

\subsection{On a determinant form with more general parameters} 
It is natural to look for a determinant expression for the DWPF 
with less restrictions on the rapidities than in Equation 
{\bf \ref{restrictions}}. We were unable to find any such
expression, even for the simplest variations on the conditions 
of Equation {\bf \ref{restrictions}}, such as allowing only one 
rapidity, such as $u_1$, to be free, and so forth. This, of 
course, is not a proof that no such generalization exists, 
but only that, if there is one, it is unlikely to be of the 
Izergin form of Equation {\bf \ref{determinant-form}}.

\subsection{The homogeneous limit}

In the homogeneous limit 
$\alpha_i \rightarrow \alpha$, and 
$\beta_j \rightarrow \beta$, a standard procedure gives

\begin{multline}
Z^{N\times N}_{TF, res, hom} = \\
\frac{
(-1)^{
\frac{ N (N-1)}{2}
}
}
{\ll \prod^{N-1}_{n=1} n! \rr^2
}
\ll 
(\alpha - \beta)(1-\alpha \beta) 
\rr^{N^2}
\ll
\sqrt{1 - \alpha^2} \sqrt{1 - \beta^2}
\rr^{N} 
\times
\\
\det 
\ll 
{\ll \frac{\partial}{\partial \alpha} \rr}^{i-1}
{\ll \frac{\partial}{\partial  \beta} \rr}^{j-1}
\ll \frac{1}{(\alpha - \beta)(1 - \alpha \beta)} \rr
\rr_{1 \leq i, j \leq N}
\label{homogeneous}
\end{multline}

\subsection{2-Toda $\tau$-function}

Because the determinant in Equation {\bf \ref{homogeneous}} is 
bi-Wronskian, with partial derivatives in {\it two} complex 
variables, it is straightforward to show \cite{hirota-book}, 
using the Jacobi identity for determinants, that it is 
a $\tau$-function of the 2-Toda partial differential equation

\begin{equation}
\frac{\partial^2}{\partial \alpha \partial \beta} 
\log(\tau_N) = 
\frac{\tau_{N+1} \tau_{N-1}}{\tau^2_N}
\end{equation}

As mentioned earlier, the homogeneous limit of Izergin's determinant 
expression of $Z^{N\times N}_{6V}$ is proportional to a bi-Wronskian 
with partial derivatives in {\it one} complex variable, and therefore 
is a $\tau$-function of the 1-Toda partial differential equation 
\cite{k-z}. 

This observation was used in \cite{k-z} to study the free energy of the 
six vertex model in the presence of DWBC. Since $Z^{N\times N}_{TF}$ 
can be computed explicitly in product form using the algebraic Bethe 
ansatz, as we will see below, the free energy can also be computed 
explicitly, and the relationship with 2-Toda remains a curious 
observation. 

\subsection{On enumeration} As is well known, $Z^{N\times N}_{6V}$ 
can be used to enumerate alternating sign matrices (ASM's) 
\cite{kuperberg-1, kuperberg-2}. At the free fermion point, 
$Z^{N\times N}_{6V}$ 2-enumerates ASM's \cite{kuperberg-2}.

$Z^{N\times N}_{TF}$ can also be used to enumerate ASM's, but because 
the model is yet again a free fermion model, one can easily show that 
here too one obtains 2-enumerations.

\subsection{A product form for $Z^{N\times N}_{TF, res}$}
The determinant in Equation {\bf \ref{determinant-form}} 
factorizes 

\begin{multline}
\det \ll \mathcal{M}^{N\times N} \rr
= 
\ll
\prod_{i < j} (1 - \alpha_i \alpha_j)
              (1 -  \beta_i  \beta_j)
\rr 
\frac{
\prod_{1 \leq i < j \leq N} (\alpha_i -  \alpha_j) 
                            ( \beta_j  -  \beta_i)
}
{
\prod_{1 \leq i, j \leq N} 
( 1       -  \alpha_i \beta_j) 
(\alpha_i -  \beta_j         )
} 
\label{special-factorized}
\end{multline}

The simplest way to see this is to notice that there is a change of 
variables that allows one to re-write the determinant in Equation 
{\bf \ref{determinant-form}} in Cauchy form\footnote{This remark is 
due to Ch Krattenthaler.}.

This is reminiscent of the factorization of Izergin's determinant 
in the six vertex model, at the free fermion point \cite{bpz}. We 
attribute the factorization of Equation {\bf \ref{special-factorized}} 
to the fact that the trigonometric Felderhof model is a free fermion 
model. 

From Equation {\bf \ref{determinant-form}} and Equation 
{\bf \ref{special-factorized}}, we obtain

\begin{equation}
Z^{N\times N}_{TF, res}
= 
\prod_{1 \leq i, j \leq N} \sqrt{1 - \alpha_{i} \alpha_{j}}
                           \sqrt{1 -  \beta_{i}  \beta_{j}} 
\label{restricted-factorized-form}
\end{equation}

The simple form of the factorized result in 
Equation {\bf \ref{restricted-factorized-form}} suggests that 
a similar result may hold in the general case with dependence 
on all parameters. This will be the topic of the next section.

\section{Product form for general $Z^{N\times N}_{TF}$}

Unlike $Z^{N\times N}_{TF, res}$, $Z^{N\times N}_{TF}$ is not 
invariant under permuting adjacent variables. All expressions 
in this section are valid only for the ordering shown 
in Figure {\bf \ref{lattice}}. Expressions of $Z^{N\times N}_{TF}$ 
with different orderings are related by factors of vertex weights.  

\subsection{Definitions} Consider weights which depend only on 
the vertical variables $\{\beta\}$ and $\{v\}$

\begin{alignat}{3}
 \check{a}_{1, ij}  &=  1 - \beta_i \beta_j e^{v_i - v_j}, 
&\check{a}_{2, ij}   =  e^{v_i - v_j} - \beta_i \beta_j 
\no \\
 \check{b}_{1, ij}  &=  \beta_i- \beta_j e^{v_i - v_j}, 
&\check{b}_{2, ij}   =  \beta_j- \beta_i e^{v_i - v_j} 
\no \\
\check{c}_{1, ij}  &= \sqrt{1 - \beta_i^{2}} \sqrt{1 - \beta_j^{2}}
	              \, \, e^{v_i - v_j} ,
\no \\
\check{c}_{2, ij}  &= \sqrt{1 - \beta_i^{2}} \sqrt{1 - \beta_j^{2}}
\end{alignat}

The $R$-matrix for the trigonometric Felderhof model is

\begin{equation*}
R_{ij}(\beta_i, \beta_j, v_i, v_j) =
\ll
\begin{array}{cccc}
\check{a}_{1, ij} & 0                 & 0                 & 0 \\
0                 & \check{b}_{1, ij} & \check{c}_{1, ij} & 0 \\
0                 & \check{c}_{2, ij} & \check{b}_{2, ij} & 0 \\
0                 & 0                 & 0                 & \check{a}_{2, ij}
\end{array}
\rr_{ij}
\end{equation*}

The monodromy matrix, $T_{0,1\ldots N}$, for $N$ sites, is

\begin{eqnarray}
T_{0, 1 \ldots N}(\alpha, u)
&=&
R_{0N}(\alpha, \beta_N, u, v_N) \ldots R_{01}(\alpha, \beta_1, u, v_1)
\no \\
&=&
\ll
\begin{array}{cc}
A_{1 \ldots N} (\alpha, u) & B_{1 \ldots N} (\alpha, u) \\
C_{1 \ldots N} (\alpha, u) & D_{1 \ldots N} (\alpha, u)
\end{array}
\rr_0
\end{eqnarray}

As explained in \cite{korepin-book}, $Z_{TF}^{N\times N}$ can be
expressed in terms of the creation operators, $B_{1\ldots N}$, as

\begin{equation}
Z_{TF}^{N\times N}
=
\langle 1|
B_{1 \ldots N}(\alpha_1, u_1)
\ldots
B_{1 \ldots N}(\alpha_N, u_N)
|0 \rangle
\label{product-1}
\end{equation}

\noindent where

$\langle 1| = \bigotimes_{j=1}^{N}
\ll
\begin{array}{cc}0 & 1
\end{array}
\rr_j
=\bigotimes_{j=1}^N \ll \downarrow_j\rr$,
and
$ |0\rangle = \bigotimes_{j=1}^{N}
\ll
\begin{array}{c}1 \\0
\end{array}
\rr_j
=\bigotimes_{j=1}^N \ll \uparrow_j\rr$.

The expression in Equation {\bf \ref{product-1}} is not easy to evaluate
directly, since the creation operators are sums containing $2^N$ terms,
and each term is a tensor product acting in all of the spaces $1,\ldots, N$.

\subsection{A factorizing matrix} Following \cite{ms}, we define
an initial {\it factorizing} $F$-matrix,
$F_{1, 2 \ldots N}$, by

\begin{eqnarray}
F_{1, 2 \ldots N}(\beta_1; \beta_2, \ldots, \beta_N; v_1; v_2, \ldots, v_N)
&=&
e^{(11)}_1+e^{(22)}_1T_{1,2\ldots N}(\beta_1,v_1) \no \\
&=&
\ll
\begin{array}{cc}\mathbf{1} & \mathbf{0} \\
C_{2\ldots N}(\beta_1,v_1)& D_{2\ldots N}
(\beta_1, v_1)
\end{array}
\rr_1
\end{eqnarray}

From that, the full $F$-matrix, $F_{1\ldots N}$, is defined 
recursively by

\begin{multline}
F_{1 \ldots N}(\beta_1, \ldots, \beta_N;v_1,\ldots,v_N)
=
F_{2 \ldots N} F_{1, 2 \ldots N}
=\\
\ldots
=
F_{(N-1)N} F_{(N-2), (N-1)N} \ldots F_{1,2 \ldots N}
\label{fullF}
\end{multline}

\noindent where

\begin{multline}
F_{(N-1)N}(\beta_{(N-1)},\beta_N;v_{(N-1)},v_N) =
\\
\ll
\begin{array}{cccc}
1 & 0        & 0        & 0        \\
0 & 1        & 0        & 0        \\
0 & \check{c}_{2, (N-1)N}   & \check{b}_{2, (N-1)N} & 0        \\
0 & 0        & 0        & \check{a}_{2, (N-1)N}
\end{array}
\rr_{(N-1)N}
\end{multline}

\subsection{A twisted monodromy matrix} The full $F$-matrix,
$F_{1\ldots N}$, is now used to construct a {\it twisted}
monodromy matrix

\begin{multline}
\widetilde{T}_{0,1\ldots N}
\ll
\alpha;\beta_1,\ldots,\beta_N;u;v_1,\ldots,v_N
\rr
=
\\
F_{1\ldots N}
T_{0,1\ldots N}
\ll
\alpha;\beta_1,\ldots,\beta_N;u;v_1,\ldots,v_N
\rr
F_{1\ldots N}^{-1}
\end{multline}

\subsection{Twisted creation operators}
We define twisted versions of the Bethe ansatz operators
as follows. 
$\widetilde{A}_{1\ldots N}(\alpha,u)
=
F_{1\ldots N}
A_{1\ldots N}(\alpha,u)
F_{1\ldots N}^{-1}$,
{\it etc}. Using the notation 
$a_{1, 0j} = 1 - \alpha \beta_j \, \, e^{u - v_j}$, 
$a_{2, 0j} = e^{u - v_j} - \alpha \beta_j$, {\it etc},
where the label $0$ indicates dependence on the horizontal variables 
$\{\alpha, u\}$, one can show that 

\begin{multline}
\widetilde{A}_{1\dots N}(\alpha,u)
=
\bigotimes_{j=1}^N
\ll
\begin{array}{cc}
{a}_{1, 0j} & 0 \\
0         & \frac{a_{2,j0}{a}_{2, 0j}} {{b}_{2, j0}}
\end{array}
\rr_j
+
\widetilde{B}_{1\ldots N}(\alpha,u)
\widetilde{D}_{1\ldots N}^{-1}(\alpha,u)
\widetilde{C}_{1\ldots N}(\alpha,u) \\ 
\label{twisted-A}
\end{multline}

\begin{multline}
\widetilde{B}_{1 \ldots N}(\alpha,u) =
\\
\sum_{l=1}^{N}
\bigotimes_{j < l}
\ll
\begin{array}{cc}
{b}_{2, 0j} & 0 \\
0         & \frac{{a}_{2, 0j} \check{a}_{2, jl}}{\check{b}_{2, jl}}
\end{array}
\rr_j
\ll
\begin{array}{cc}
0        & 0 \\
{c}_{1, 0l} & 0
\end{array}
\rr_l
\bigotimes_{j>l}
\ll
\begin{array}{cc}
\check{a}_{1, lj} {b}_{2, 0j} & 0 \\
0 &
\frac{\check{a}_{2, lj}{a}_{2, 0j}\check{a}_{2,
jl}}{\check{b}_{2, jl}}
\end{array}
\rr_j 
\label{twisted-B}
\end{multline}

\begin{multline}
\widetilde{C}_{1\ldots N}(\alpha,u)
= \\
\sum_{l=1}^{N}
\bigotimes_{j<l}
\ll
\begin{array}{cc}\frac{\check{a}_{1,lj}b_{2,0j}}{\check{b}_{2,lj}} & 0 \\
0 &a_{2,0j}
\end{array}
\rr_j
\ll
\begin{array}{cc}0 & c_{2,0l} \\
0 & 0\end{array}
\rr_l
\bigotimes_{j>l}
\ll
\begin{array}{cc}\frac{b_{2,0j}}{\check{b}_{2,lj}} & 0 \\
0 & \frac{a_{2,0j}}{\check{a}_{2,lj}}
\end{array}
\rr_j
\label{twisted-C}
\end{multline}

\begin{multline}
\widetilde{D}_{1\ldots N}(\alpha,u)
=
\bigotimes_{j=1}^{N}
\ll
\begin{array}{cc}b_{2,0j} & 0 \\
0 & a_{2,0j}
\end{array}
\rr_j \\
\label{twisted-D}
\end{multline}

\subsection{Remarks on proof of Equations 
{\bf \ref{twisted-A}, 
     \ref{twisted-B}, 
     \ref{twisted-C} and
     \ref{twisted-D}}
} One first verifies by direct
computation that the above formulas hold for 
$\widetilde{A}_{12}(\alpha,u)$,
$\widetilde{B}_{12}(\alpha,u)$, 
$\widetilde{C}_{12}(\alpha,u)$ and
$\widetilde{D}_{12}(\alpha,u)$. 
This becomes the basis for a proof by induction, in which each 
formula is proven individually. For example, to prove the formula 
for the twisted operator $\widetilde{B}_{1\ldots N}$, we observe 
that, by construction,  the untwisted operator $B_{1\ldots N}$ 
satisfies

\begin{equation}
B_{1\ldots N}(\alpha,u)
=
A_{2\ldots N}(\alpha,u)
\ll
\begin{array}{cc}
0 & 0 \\
c_{1,01} & 0
\end{array}
\rr_1
+
B_{2\ldots N}(\alpha,u)
\ll
\begin{array}{cc}
b_{2,01} & 0 \\
0 & a_{2,01}
\end{array}
\rr_1
\label{B1...N}
\end{equation}

Multiplying Equation {\bf \ref{B1...N}} from the left by $F_{1\ldots N}$,
and from the right by $F^{-1}_{2\ldots N}$, we find that 
$\widetilde{B}_{1\ldots N}(\alpha,u)$ satisfies

\begin{multline}
\widetilde{B}_{1\ldots N}(\alpha,u)
F_{1\ldots N}
F^{-1}_{2\ldots N}
= \\
F_{1\ldots N}
F^{-1}_{2\ldots N}
%\times
%\\
\left\{
\widetilde{A}_{2\ldots N}(\alpha,u)
\ll
\begin{array}{cc}
0 & 0 \\
c_{1,01} & 0
\end{array}
\rr_1
+
\widetilde{B}_{2\ldots N}(\alpha,u)
\ll
\begin{array}{cc}
b_{2,01} & 0 \\
0 & a_{2,01}
\end{array}
\rr_1
\right\}
\label{twistB1...N}
\end{multline}

From Equation {\bf \ref{fullF}}, we have
$F_{1\ldots N}
F^{-1}_{2\ldots N}
=
F_{2\ldots N}
F_{1,2\ldots N}
F_{2\ldots N}^{-1}$,
so Equation {\bf \ref{twistB1...N}} becomes the following matrix 
equation in space 1

\begin{multline}
\widetilde{B}_{1\ldots N}(\alpha,u)
\ll
\begin{array}{cc}\mathbf{1} & \mathbf{0} \\
\widetilde{C}_{2\ldots N}(\beta_1,v_1)& \widetilde{D}_{2\ldots N}
(\beta_1, v_1)
\end{array}
\rr_1
= 
\\
\ll
\begin{array}{cc}\mathbf{1} & \mathbf{0} \\
\widetilde{C}_{2\ldots N}(\beta_1,v_1)& \widetilde{D}_{2\ldots N}
(\beta_1, v_1)
\end{array}
\rr_1
\times
\\
\left\{
\widetilde{A}_{2\ldots N}(\alpha,u)
\ll
\begin{array}{cc}
0 & 0 \\
c_{1,01} & 0
\end{array}
\rr_1
+
\widetilde{B}_{2\ldots N}(\alpha,u)
\ll
\begin{array}{cc}
b_{2,01} & 0 \\
0 & a_{2,01}
\end{array}
\rr_1
\right\}
\label{matrix-equation}
\end{multline}

This is a recursion relation for 
$\widetilde{B}_{1\ldots N}(\alpha,u)$ in terms of twisted
operators over the $N-1$ spaces $2, \ldots, N$. We use it 
to prove Equation {\bf \ref{twisted-B}} inductively, as 
follows.

First, we postulate that the expressions of Equations 
{\bf 
\ref{twisted-A},
\ref{twisted-B},
\ref{twisted-C}}
and
{\bf \ref{twisted-D}}
hold over the $N-1$ spaces $2,\ldots, N$. 
Next, we substitute them into Equation {\bf \ref{matrix-equation}}. 
From that, the expression in Equation 
{\bf \ref{twisted-B}} for $\widetilde{B}_{1\ldots N}$ is seen to be 
the unique solution to Equation {\bf \ref{matrix-equation}}. 
Repeating this procedure for the other twisted operators, we 
prove the postulate over the $N$ spaces $1,\ldots, N$.

\subsection{A recursion relation} 
Since 
$\langle 1|F_{1\ldots N} = \prod_{j<k}\check{a}_{2, jk}\langle 1|$ 
and $F_{1\ldots N}^{-1}|0\rangle = |0\rangle$, we can re-write
$Z_{TF}^{N\times N}$ in terms of the twisted creation operators

\begin{equation}
Z_{TF}^{N\times N}
=
\frac{
\langle 1| \widetilde{B}_{1\ldots N}(\alpha_1, u_1) \ldots
           \widetilde{B}_{1\ldots N}
(\alpha_N, u_N)|0\rangle}
{
\prod_{j<k} \check{a}_{2,jk}
}
\label{product-2}
\end{equation}

Following \cite{kmt}, we use
the expression in Equation {\bf \ref{product-2}}, together with
the explicit expression for the twisted $B$-operator, to derive
the recursion relation

\begin{equation}
Z_{TF}^{N\times N}
=
\sum_{i=1}^{N}    c_{1,Ni}
\prod_{j=1}^{N-1}   a_{2,ji}
\prod_{k \not = i}
\frac{
b_{2,Nk}
}
{
\check{b}_{2,ik}
}
\prod_{k<i} \check{a}_{2,ik}
\prod_{k>i} \check{a}_{1,ik}
Z^{(N-1)\times (N-1)}_{TF,\ (Ni)}
\label{recursion2}
\end{equation}

\noindent where the subscripts in $Z^{(N-1)\times (N-1)}_{TF,\ (Ni)}$
indicate the omission of the variables $\{\alpha_N,u_N\}$ and
$\{\beta_i,v_i\}$.

\subsection{Remarks on proof of Equation {\bf \ref{recursion2}}} 
Acting with 
$\widetilde{B}_{1\ldots N}(\alpha_N,u_N)$ on 
$|0\rangle$, in Equation {\bf \ref{product-2}}, 
we immediately find

\begin{multline}
Z_{TF}^{N\times N}
= 
\frac{
\displaystyle\sum_{i=1}^{N}
c_{1,Ni}
\displaystyle\prod_{k\not=i}
b_{2,Nk}
\displaystyle\prod_{k>i}
\check{a}_{1,ik}
}
{
\displaystyle\prod_{j<k}
\check{a}_{2,jk}
}
\times
\\
\bigotimes_{j=1}^N
\ll \downarrow_j \rr
\widetilde{B}_{1\ldots N}(\alpha_1, u_1)
\ldots
\widetilde{B}_{1\ldots N}(\alpha_{(N-1)}, u_{(N-1)})
\ll \downarrow_i \rr
\bigotimes_{j\not=i}
\ll \uparrow_j \rr
\label{recursion3}
\end{multline}

Using the identities
$
\ll \downarrow_i \rr
\sigma^{-}_i
\ll \downarrow_i \rr
=
(\sigma^{-}_i)^2
=0
$,
one obtains 

\begin{multline}
\ll \downarrow_i \rr
\widetilde{B}_{1\ldots N}(\alpha_1, u_1)
\ldots
\widetilde{B}_{1\ldots N}(\alpha_{(N-1)}, u_{(N-1)})
\ll \downarrow_i \rr
=
\\
\ll
\prod_{j=1}^{N-1} a_{2,ji}
\rr
\ll
\prod_{k\not= i}
\frac{
\check{a}_{2,ik}
}
{
\check{b}_{2,ik}
}
\rr
\ll
\prod_{k<i} \check{a}_{2,ki}
\rr
\times 
\\
\widetilde{B}_{1\ldots (i-1)(i+1)\ldots N}(\alpha_1, u_1)
\ldots
\widetilde{B}_{1\ldots (i-1)(i+1)\ldots N}(\alpha_{(N-1)}, u_{(N-1)})
\label{step}
\end{multline}

Equation {\bf \ref{recursion2}} comes from substituting Equation 
{\bf \ref{step}} in Equation {\bf \ref{recursion3}}, and using 
the fact that

\begin{multline}
Z^{(N-1)\times (N-1)}_{TF,\ (Ni)}
=
\frac{
\prod_{j<i}
\check{a}_{2,ji}
\prod_{i<k}
\check{a}_{2,ik}
}
{
\prod_{j<k}
\check{a}_{2,jk}
}
\times
\\
\bigotimes_{j\not= i}
\ll \downarrow_j \rr
\widetilde{B}_{1\ldots (i-1)(i+1)\ldots N}(\alpha_1, u_1)
\ldots
\widetilde{B}_{1\ldots (i-1)(i+1)\ldots N}(\alpha_{(N-1)}, u_{(N-1)})
\bigotimes_{j\not= i}
\ll \uparrow_j \rr
\end{multline}

%The recursion relation in Equation {\bf \ref{recursion2}}
%has a simple graphical interpretation. It is basically 
%the statement that there is one and only one $c_1$ vertex 
%on the bottom row of the DWBC lattice. This allows one
%to split $Z^{N\times N}_{TF}$ into $N$ parts. Using the 
%the algebraic Bethe ansatz commutation relations, or 
%equivalently, the Yang-Baxter equations, each of these 
%parts can be re-written as one of the terms in Equation 
%{\bf \ref{recursion2}}. A similar exercise was carried
%out in the context of the six vertex model in \cite{bpz, fp}.

\subsection{Solution of recursion equation} It can be shown 
that the following product expression satisfies Equation 
{\bf \ref{recursion2}}

\bigskip
\begin{boxedminipage}[c]{12.5cm}
\begin{multline}
Z_{TF}^{N \times N}
=
\\
\ll
\prod_{k=1}^{N}
e^{k(u_k - v_k)}
\sqrt{1 - \alpha_k^{2}}
\sqrt{1 -  \beta_k^{2}}
\rr
\ll
\prod_{1 \leq j < k \leq N}
(e^{u_j-u_k}-\alpha_j\alpha_k)
(e^{v_k-v_j}-\beta_k \beta_j)
\rr
\\
=
\ll
\prod_{k=1}^{N}
e^{k(u_k - v_k)}
c_{2}(\alpha_k,\beta_k)
\rr
\ll
\prod_{1 \leq j < k \leq N}
{a}_2(\alpha_j,\alpha_k,u_j,u_k)
{a}_2(\beta_k,\beta_j,v_k,v_j)
\rr
\label{PF}
\end{multline}
\end{boxedminipage}
\bigskip 

\subsection{Remarks on proof of Equation {\bf \ref{PF}}} 
From Equation {\bf \ref{PF}}, we have

\begin{multline}
Z^{(N-1)\times (N-1)}_{TF,\ (Ni)}
=
\\
\frac{
\ll
e^{-Nu_N+iv_i}
\displaystyle\prod_{k=i+1}^{N}
e^{v_k}
\rr
}
{
\ll
\displaystyle\prod_{j=1}^{N-1}
{a}_2(\alpha_j,\alpha_N,u_j,u_N)
\rr
\ll
\displaystyle\prod_{j < i}
{a}_2(\beta_i, \beta_j, v_i, v_j)
\displaystyle\prod_{i < j}
{a}_2(\beta_j, \beta_i, v_j, v_i)
\rr
} 
\times
\\
\frac{1}{c_2(\alpha_N,\beta_i)}
Z_{TF}^{N\times N}
\label{recursion4}
\end{multline}

Using Equation {\bf \ref{recursion4}}, and after considerable 
manipulation, one recovers

\begin{multline}
\sum_{i=1}^{N}    c_{1,Ni}
\prod_{j=1}^{N-1}   a_{2,ji}
\prod_{k \not = i}
\frac{
b_{2,Nk}
}
{
\check{b}_{2,ik}
}
\prod_{k<i} \check{a}_{2,ik}
\prod_{k>i} \check{a}_{1,ik}
Z^{(N-1)\times (N-1)}_{TF,\ (Ni)}
=
\\
\sum_{i=1}^{N}
\ll
\prod_{j=1}^{N-1}
\frac{
a_2(\alpha_j,\beta_i,u_j,v_i)
}
{
a_2(\alpha_j,\alpha_N,u_j,u_N)
}
\rr
\ll
\prod_{k\not= i}
\frac{
b_2(\beta_k,\alpha_N,v_k,u_N)
}
{
b_2(\beta_k,\beta_i,v_k,v_i)
}
\rr
Z_{TF}^{N\times N}
\label{recursion5}
\end{multline}

Finally, we observe that

\begin{multline}
\sum_{i=1}^{N}
\ll
\prod_{j=1}^{N-1}
a_2(\alpha_j,\beta_i,u_j,v_i)
\rr
\ll
\prod_{k\not= i}
b_2(\beta_k,\alpha_N,v_k,u_N)
\rr
\ll
\prod_{\substack{j\not= k \\ k\not= i }}
b_2(\beta_j,\beta_k,v_j,v_k)
\rr
=
\\
\ll
\prod_{j=1}^{N-1}
a_2(\alpha_j,\alpha_N,u_j,u_N)
\rr
\ll
\prod_{j\not= k}
b_2(\beta_j,\beta_k,v_j,v_k)
\rr
\label{product-form}
\end{multline}
which can be checked by noticing that both sides are polynomials of 
degree $N-1$ in the variable $\alpha_N$, and furthermore the equality 
is satisfied at the $N$ points $\alpha_N = \beta_j e^{v_j-u_N}$, where
$j = 1, \ldots, N$. Equation {\bf\ref{product-form}} means that

\begin{equation}
\sum_{i=1}^{N}
\ll
\prod_{j=1}^{N-1}
\frac{
a_2(\alpha_j,\beta_i,u_j,v_i)
}
{
a_2(\alpha_j,\alpha_N,u_j,u_N)
}
\rr
\ll
\prod_{k\not= i}
\frac{
b_2(\beta_k,\alpha_N,v_k,u_N)
}
{
b_2(\beta_k,\beta_i,v_k,v_i)
}
\rr
=
1
\end{equation}

\noindent which, when substituted in Equation {\bf\ref{recursion5}}, 
causes it to collapse to the recursion relation Equation 
{\bf\ref{recursion2}}, as required.

\section*{Acknowledgements}

We thank T~Deguchi, N~Kitanine and Ch~Krattenthaler for useful 
comments. AC, MW and MZ are supported by Australian Postgraduate 
Awards.

\end{document}